\documentclass[twocolumn,showpacs,aps,prl]{revtex4}

\usepackage{amsmath,amsfonts,amssymb}
\usepackage{wrapfig}
\usepackage{graphicx}
\usepackage{bbm}
\usepackage{color}

\newcommand{\ket}[1]{| #1 \rangle}
\newcommand{\bra}[1]{\langle #1 |}

\newcommand{\ex}[1]{\langle #1 \rangle}

\newcommand{\beq}{\begin{eqnarray}}
\newcommand{\eeq}{\end{eqnarray}}

\begin{document}

\title{Witnessing Quantum Coherence: from solid-state to biological systems}
\author{Che-Ming Li$^{1,2,3}$, Neill Lambert$^{3}$, Yueh-Nan Chen$^{4}$,
Guang-Yin Chen$^{4}$, and Franco Nori$^{3,5}$}
\affiliation{$^{1}$Department of Engineering Science and Supercomputing Research Center, National Cheng Kung University, Tainan 70101, Taiwan}
\affiliation{$^{2}$Supercomputing Research Center, National Cheng Kung University, Tainan 701, Taiwan}
\affiliation{$^{3}$Advanced Science Institute, RIKEN, Saitama 351-0198, Japan}
\affiliation{$^{4}$Department of Physics and National Center for Theoretical Sciences, National Cheng Kung
University, Tainan 701, Taiwan}
\affiliation{$^{5}$Department of Physics, University of Michigan, Ann Arbor, Michigan 48109-1040 USA}

\begin{abstract}
Quantum coherence is one of the primary non-classical features of
quantum systems. While protocols such as the Leggett-Garg
inequality (LGI) and quantum tomography can be used to test for
the existence of quantum coherence and dynamics in a given system,
{\em unambiguously} detecting inherent ``quantumness'' still faces
serious obstacles in terms of experimental feasibility and
efficiency, particularly in complex systems. Here we introduce two
``quantum witnesses'' to efficiently verify quantum coherence and
dynamics in the time domain, without the expense and burden of
non-invasive measurements or full tomographic processes. Using
several physical examples, including quantum transport in
solid-state nanostructures and in biological organisms, we show
that these quantum witnesses are robust and have a much finer
resolution in their detection window than the LGI has. These
robust quantum indicators may assist in reducing the experimental
overhead in unambiguously verifying quantum coherence in complex
systems.
\end{abstract}

\maketitle

\section*{Introduction}

Quantum coherence, or superposition, between different states is one of the main features of quantum
systems. This distinctive property, coherence, ultimately leads to
a variety of other phenomena, e.g.,
entanglement \cite{Amico2008,Horodecki09}.  It is also thought to
be the power behind several ``quantum tools'', including quantum
information processing\cite{Pan08}, metrology \cite{Giovannetti04},
transport \cite{Brandes05}, and recently, some functions in
biological organisms \cite{Cheng09} (e.g., efficient energy
transport).

Identifying quantum coherence and dynamics in an efficient way,
given limited system access, is indispensable for ensuring
reliable quantum applications in a variety of contexts.
Furthermore, the question of whether quantum coherence can really
exist in biological organisms {\em in vivo}, e.g., in a
photosynthetic complex or in an avian chemical compass, surrounded
by a hot and wet environment, has triggered a surge of interest
into the relationship between quantum coherence and biological
function \cite{Ishizaki09,Ishizaki10}.  In these cases, full-system
access is often very limited, and signatures of quantum coherence
are often indirect.

The existing methods for identifying quantum coherent behavior can
be generally classified into two types.  The first type are based
on imposing what can be thought of as a classical constraint \cite{adam}, such
as macroscopic realism and non-invasive measurements in the Leggett-Garg
inequality (LGI) \cite{Leggett85}, or realism and locality in
Bell's Inequality. Even though inequalities like the LGI were
originally envisaged as a fundamental test of physical theories, a
violation of the LGI can also be considered as a tool for
classifying the behavior observed in experiment as quantum or
classical. However, the Leggett-Garg inequality faces severe
experimental difficulties when used as such a tool as it requires
\emph{noninvasive} measurements, e.g., via quantum nondemolition
(QND) measurement \cite{Caves80,Bocko96}, weak
measurement \cite{Aharonov90}, or quantum-gate-assisted ideal
non-invasive measurements \cite{Knee12}. Because of this only a
few tests of the LGI have been
reported \cite{Knee12,PL10,Dressel11,Goggin11,Waldherr11}.

The second type of test is based on deduction; do the results of a
given experiment sufficiently correspond to the predictions of
quantum theory (or classical theory, depending on the approach).
Quantum witnesses can be considered as one such test, as they use
the knowledge of a quantum state or of some quantum dynamics to
determine whether an experimental system possesses quantum
properties. Some examples that have been employed elsewhere
include witnesses of entanglement \cite{Guhne09,jian}, direct
measurement of coherence terms of density matrices, or the
analysis of process tomography \cite{Nielsen00} for non-classical
state evolutions. The experimental realization of this kind of
verification usually needs tomographic techniques, and then the
required experimental resources in terms of measurement settings
increases exponentially with the system
complexity \cite{Guhne09,Nielsen00,Li10}. Moreover, quantum state
and process tomography are still difficult to implement in general
systems and for general state evolutions, e.g., particularly in
systems like charge transport through nanostructures, the transfer
of electronic excitations in a photosynthetic complex, or systems
where the state space is large.

In this work, we introduce two quantum witnesses to verify
quantum coherence and dynamics in the time domain, both of which
have various advantages and disadvantages. Both are efficient in the
sense that there is \emph{no} need to perform noninvasive
measurements or to use quantum tomography, dramatically reducing
the overhead and complexity of unambiguous experimental
verification of quantum phenomena.

We apply these quantum witnesses to five examples: (1)
electron-pair tunnelling in a Cooper-pair box and coherent
evolution of single-transmon qubit, (2) charge transport through
double quantum dots, (3) non-equilibrium energy transfer in the
photosynthetic pigment-protein complex, (4) vacuum Rabi
oscillation in lossy cavities, and (5) coherent rotations of
photonic qubits. Furthermore, as we will illustrate in these
examples, our quantum witnesses possess a finer detection
resolution than the LGI.

Both witnesses, which we will introduce shortly, involve the
following steps: (Figure 1a): first, we prepare the system in a
known product state with its environment (or reservoir, here we
use both terms interchangeably) $\rho_{S\!R}(0)$. We then let
$\rho_{S\!R}(0)$ evolve for a period of time $t_{0}$, to reach the
state $\rho_{S\!R}(t_{0})$ (during which time one hopes the state
has acquired significant coherence due to its internal dynamics).
The second step is to implement a quantum witness using a
``correlation check'' between the state $\rho_{S\!R}(t_{0})$ and
its state at another later time $t\geq t_0$, $\rho_{S\!R}(t)$. The
goal of this correlation check is to investigate non-classical
properties in these two-time state-state correlations (see Figure
1b). If the state $\rho_{S\!R}(t_{0})$ can be detected then by
our quantum witness as having quantum properties, this implies
that either the system state
$\rho_{S}(t_{0})=\text{Tr}_{R}[\rho_{S\!R}(t_{0})]$ possesses
significant quantum coherence or that the state
$\rho_{S\!R}(t_{0})$ is an entangled system-bath state.

\begin{figure}
\includegraphics[width=8.5cm]{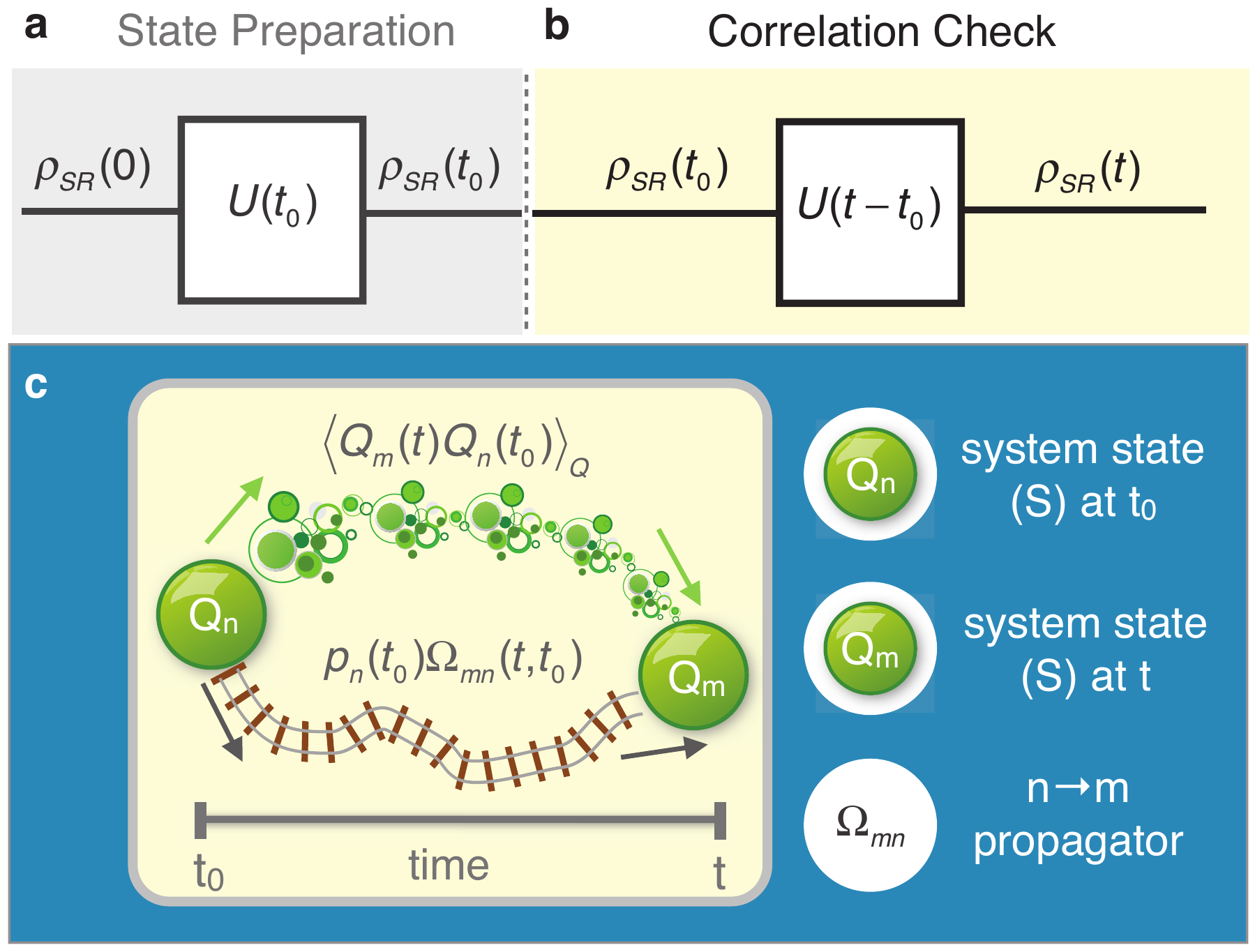}
\caption{\textbf{Detecting quantum coherence and dynamics.}
Generic procedure for detecting quantum coherence: We need
\textbf{(a)} state preparation and \textbf{(b)} a correlation
check. With the freedom to manipulate the system state, the
initial state of the total system can be reasonably prepared as a
product state $\rho_{S\!R}(0)=\rho_{S}(0)\otimes \rho_{R}(0)$,
where $\rho_{R}(0)$ is the reservoir state. \textbf{(b)} and
\textbf{(c)} show the correlation check, or measurements, we base
our quantum witnesses on. Assuming the system state at time
$t_{0}$ is $\rho_{S}(t_{0})=Q_{n}(t_{0}=p_{n}(t_{0}))$, the
probability of being in a state $Q_{m}$ at later time $t$ is
determined by the $n\rightarrow m$ propagator
$\Omega_{mn}(t,t_{0})$. The general quantum correlator is defined by
$\left \langle Q_{m}(t)Q_{n}(t_{0})\right\rangle_{Q}$ (upper green
path) and the classical one is defined by
$p_{n}(t_{0})\Omega_{mn}(t,t_{0})$ (lower brown track). In this
expression, $\Omega_{mn}(t,t_0)$ is the probability of measuring
the system in state $m$ at time $t$ given that it was in the state
$n$ at time $t_0$.  Both definitions describe the connections
between $Q_{n}(t_{0})$ and $Q_{m}(t)$ for arbitrary states
$\rho_{S}(t_{0})$ with a distribution of state populations
$\{p_{n}(t_{0})\}$. As shown in the Methods, all classical
dynamics should satisfy the relation, $\left \langle
Q_{m}(t)Q_{n}(t_{0})\right\rangle_{Q} =
p_{n}(t_{0})\Omega_{mn}(t,t_{0})$, whereas violations of the
equality reveal a signature of quantum dynamics. While our quantum
witnesses are derived from this ``correlation check'', the
experimental requirements for each witness differ.}
\end{figure}

\section*{Results}

In order to find a signature of quantum dynamics we start by seeking characteristic
features of classical dynamics or states \cite{vedral}. All
separable mixtures of system-reservoir states, with no coherent
components, which we call \emph{classical states}, obey the
following relation for their two-time correlations:
\begin{equation}
\left \langle
Q_{m}(t)Q_{n}(t_{0})\right\rangle_{Q}=p_{n}(t_{0})\Omega_{mn}(t,t_{0}).\label{c1}
\end{equation}
See Methods for the proof. Succinctly put, equation (\ref{c1})
implies it is possible to define all future behavior based on only
the system's instantaneous expectation
values $p_n(t_0)$. However, most quantum
correlation functions also obey this relation under certain measurement conditions. For example, a correlation
function constructed from two-time projective measurements has this form as the measurement
at $t_0$ destroys the coherence in the state at that time.  Here
$Q_i$ is an observable which measures if the system is in the state $i$.  This state is assumed to have a
classical meaning (e.g., localized charge state, etc) and the observable is
normalized so that its expectation value is directly equal to the
probability of observing the system in that state $\left\langle Q_i\right\rangle =
p_i$. The propagator $\Omega_{mn}(t,t_0)$ is the probability of
measuring the system in state $m$ at time $t$ given that it was in
the state $n$ at time $t_0$  (and which in principle depends on
the state of the reservoir, so can include classical non-Markovian
correlations, see Methods). Several other recent tests of quantumness \cite{Waldherr11,Huelga95,Lambert10,Lambert10b,Lambert11} rely on imposing
Markovianity on $\Omega_{nm}(t,t_0)$.  In our first witness we avoid taking that approach so that we can still distinguish quantum from classical non-Markovian dynamics.  However we will use it in our second witness.

In principle, one could use Eq. (\ref{c1}) to construct a quantum witness of the form:
\begin{equation}
\mathcal{W}_{QQ}:=\bigg|\left \langle Q_{m}(t)Q_{n}(t_{0})\right\rangle_{Q}-p_{n}(t_{0})\Omega_{mn}(t,t_{0})\bigg|.\label{wqq}
\end{equation}
Where a non-zero result $\mathcal{W}_{QQ}>0$, implies the state
at $t_0$ can be considered as quantum in that it contains quantum coherence which effect its future evolution.
  However, as mentioned above, most quantum correlation functions
also obey equation (\ref{c1}), which will give $\mathcal{W}_{QQ}=0$.  Is it ever possible to
observe a non-zero $\mathcal{W}_{QQ}$?  In some
 cases coherence, or ``amplitude'', sensitive correlation functions
are encountered in quantum optics~\cite{Gardiner}, and in linear-response theory~\cite{Mukamel95}. However,
 these are typically extracted from spectral functions in the steady state, or put in a symmetrized form, in which case
 any affect on the correlation function from the initial state coherence may be lost.  In all the examples we consider in this work
this witness $\mathcal{W}_{QQ}$ cannot be directly measured, as
the initial coherence is of course destroyed by the first (projective) measurement.
Fortunately, $\mathcal{W}_{QQ}$, via Eq. (\ref{c1}), gives us a way to develop a
more generally applicable and valid witness.

\subsection{Witness 1}

Our first practical witness (which is the main result of this work) can be
derived from Eq. (\ref{c1}) by including normalization.  Noting
that all classical system-reservoir states obey,
\begin{eqnarray}
\left \langle Q_{m}(t)\right\rangle=\sum_{n=1}^{d}p_{n}(t_{0})\Omega_{mn}(t,t_{0}),\label{c2}
\end{eqnarray}
where $d$ is the number of states $n$ in, or dimensionality
of, the system state space, we define our first quantum witness
as
\begin{equation}
\mathcal{W}_{Q}:=\bigg|\left \langle Q_{m}(t)\right\rangle-\sum_{n}p_{n}(t_{0})\Omega_{mn}(t,t_{0})\bigg|.\label{wq}
\end{equation}
If $\mathcal{W}_{Q}>0$, we can define the
state at $t_0$ as quantum. Compared with the witness
$\mathcal{W}_{QQ}$ and the tests of the LGI, $\mathcal{W}_{Q}$ can
always be directly measured, and ideal non-invasive measurements
are not necessary. In experimental realizations, measuring the
population-related quantities, or expectation values, $\left
\langle Q_{m}(t)\right\rangle$ and $\{p_{n}(t_{0})\}$, is
generally more feasible than constructing full correlation
functions, particularly in systems which rely on destructive
(e.g., fluorescence) measurements.  Where correlation functions
can be measured with projective measurements, the second term can
of course be replaced with
$\sum_{n}p_{n}(t_{0})\Omega_{mn}(t,t_{0})\equiv \sum_n
\ex{Q_m(t)Q_n(t_0)}$.

However, determining all the propagators $\Omega_{mn}(t,t_{0})$ with
which to construct the witness requires, in principle, that we can
prepare the system in each one of it states $n$ exactly
(or, alternatively if correlation functions constructed from
projective measurements are available, it requires that we measure
every possible cross-correlation $\sum_n \ex{Q_m(t)Q_n(t_0)}$). In
the former case (where we use state preparation) we trade-off the
need to do non-invasive state measurement with the need to perform
ideal state preparation.  In complex systems it may be difficult
to prepare the system in each one of its states to construct these
propagators, and in some cases we may not even have knowledge of
the full state-space of the system.

Importantly, this problem can be easily overcome by noticing that
the individual terms in the sum in Eq.~(\ref{wq}) are always
positive. Thus when constructing the sum we can stop as soon as
the witness is violated by this partial summation (i.e., when the
terms in the summation together are larger than $\left \langle
Q_{m}(t)\right\rangle$), reducing the experimental overhead
substantially (see Figure 4 for a practical example, where we show
it is sufficient to include just one term in the sum of
Eq.~(\ref{wq})).

Note that with this witness we do not distinguish between just
system-coherence or quantum correlations (entanglement) between
system and bath/reservoir (see Methods). In addition, if there are
classical correlations between system and reservoir, i.e.,
classical non-Markovian effects \cite{Breuer07}, then some
additional experimental overhead is needed to eliminate this from
giving a ``false positive''. If this overhead is ignored this
represents a ``loop-hole'' in this witness, and in some situations
may be an obstacle for its unambiguous application. We will
discuss this explicitly later with an example of a photosynthetic
light-harvesting complex where the system and reservoir are
strongly correlated both classically and quantum mechanically.

\subsection{Witness 2} For our second witness we impose the extra condition that
$\Omega_{mn}(t,t_0)=\Omega_{mn}(t',t_0')$ for $t-t_0=t'-t_0'=\tau$, for
any time interval $\tau$.  This assumption restricts us to a
widely-studied subset of quantum processes where the system-bath/reservoir interaction is Markovian. We will show that,
under the assumption that our system lies within this subset,
quantum properties can be identified without needing to explicitly
measure propagators (i.e., neither exact state initialization or
non-invasive measurements are required). The trade-off in this
case is that the witness cannot distinguish certain types of
classical dynamics (e.g., classical non-Markovian), from quantum
properties of the system. Still, this witness exceeds the tests
proposed in earlier works under the same constraints which still
required either non-invasive measurements or state
preparation \cite{Waldherr11, Lambert10}.

This subset of quantum processes can be described as
having weak coupling between system and reservoir so that
system-reservoir state is always a product state, and the bath/reservoir
state does not evolve in time, i.e., $\rho_{R}(t)=\rho_{R}(0)$. A
large number of systems exist in this regime \cite{Breuer07}, with
well-developed models such as the master equation under the Born
approximation operating within this class (see, e.g., \cite{Breuer07,Carmichael03,Schlosshauer10}). For such cases, we
can extend the first witness so that we replace the need to
prepare the system state with that of needing to repeatedly
measure expectation values (not correlation functions) a number of
times that scales linearly with system size. To show this, we
consider an extension of Eq.~(\ref{c2}) involving a system of $d$
linear equations represented in matrix multiplication form as
follows:
\begin{equation}
\bold{P}_{j}\bold{\Omega}_{mj}=\bold{Q}_{mj}
\end{equation}
where the $d\times d$ matrix $\bold{P}_{j}$ has elements
$[\bold{P}_{j}]_{kn}=p_{n}(t_{0[j,k]})$, and $\bold{\Omega}_{mj}$
and $\bold{Q}_{mj}$ are $d\times 1$ column vectors with elements
$[\bold{\Omega}_{mj}]_{n1}=\Omega_{mn[j]}(\tau)$ and
$[\bold{Q}_{mj}]_{k1}=\left \langle
Q_{m}(t_{[j,k]})\right\rangle$, respectively. Here, $t_{0[j,k]}$
and $t_{[j,k]}$ constitute the $j$th nontrivial time-domain set
$T_{j}$:$\{t_{0[j,k]},t_{[j,k]} |\
t_{[j,k]}-t_{0[j,k]}=\tau;k=1,2,...,d\}$. For a given time
difference $\tau$ and a time pair $(t_{0[j,k]},t_{[j,k]})\in
T_{j}$, one can use the most experimentally-feasible method of
measurement, i.e., \emph{invasive measurement}, to obtain the
information about the state populations, $p_{n}(t_{0[j,k]})$ and
the expectation values $\left \langle
Q_{m}(t_{[j,k]})\right\rangle$.

\begin{figure}[th]
\includegraphics[width=8.7cm]{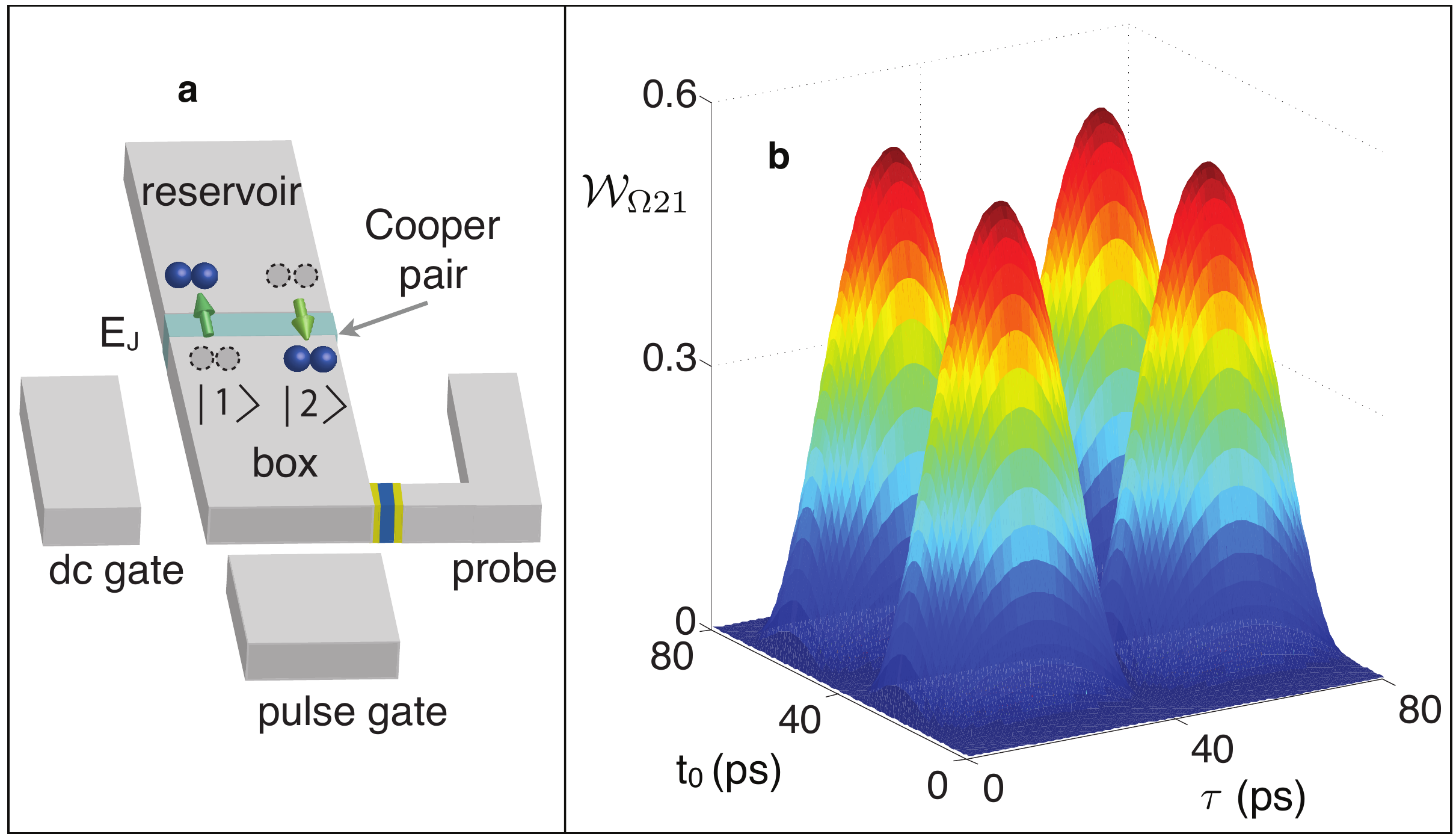}
\caption{\textbf{Detecting quantum oscillations.} \textbf{(a)}
shows a schematic circuit of a single-Cooper-pair box
\cite{you,Nakamura99}. Its Hamiltonian is described by
$H_{C\!B}=E_{C}(1-2n_{g})/2(\left|2\right\rangle\!\!\left\langle
2\right|-\left|1\right\rangle\!\!\left\langle
1\right|)-E_{J}/2(\left|1\right\rangle\!\!\left\langle
2\right|+\left|2\right\rangle\!\!\left\langle 1\right|)$, where
$E_{J}$ and $E_{C}$ are the Josephson energy and the
single-Cooper-pair charging energy of the box, respectively. The
relative energy of the state with no excess Cooper pairs  in the
box $\left|1\right\rangle$ and the state with one excess Cooper
pair $\left|2\right\rangle$, is controlled through the gate
voltage, which is parametrized by $n_{g}$. The resonance of the
states $\left|1\right\rangle$ and $\left|2\right\rangle$ can be
brought from the initial state $\left|1\right\rangle$ by the
applied voltage pulse for $n_{g}=0.5$. \textbf{(b)}, Detecting
quantum dynamics in the resonance of the two charge states
($n_{g}=0.5$) with the quantum witness $\mathcal{W}_{\Omega 21}$.
We use the realistic parameter $E_{J}=51.8$ $\mu$eV. The positive regions are identified as
the quantum areas. 
}\end{figure}

Given a set of measurement results to sufficiently describe the
state populations, the vector $\bold{\Omega}_{mj}$ can be
determined by simple algebraic methods. For nonzero determinant
$\text{det}(\bold{P}_{j})$, we have
$\Omega_{mn[j]}(\tau)=\text{det}(\bold{P}_{mj}^{(n)})/\text{det}(\bold{P}_{j})$,
where $\bold{P}_{mj}^{(n)}$ is the matrix formed by replacing the
$n$th column of $\bold{P}_{j}$ by $\bold{Q}_{mj}$. For an
arbitrary pair of time-domain sets, say $T_{j}$ and $T_{j'}$, we
impose an additional condition (not used in the earlier witnesses)
that their propagators should be identical for all classical
systems (within the subset described above):
$\bold{\Omega}_{mj}=\bold{\Omega}_{mj'}$.  If the system and its
environment are classically-correlated, i.e., they are not in a
product state, this assumption does not hold.  Any comparison
between $\bold{\Omega}_{mj}$ and $\bold{\Omega}_{mj'}$ can be
considered as a quantum witness for this subset, such as the
vector-element comparison:
 \begin{equation}
\mathcal{W}_{\Omega mn}:=\bigg|\text{det}(\bold{P}_{mj}^{(n)})\text{det}(\bold{P}_{j'})-\text{det}(\bold{P}_{mj'}^{(n)})\text{det}(\bold{P}_{j})\bigg|.\label{wo}
\end{equation}
If $\mathcal{W}_{\Omega mn}>0$, and under the assumptions
described earlier, we can again assume that some of the (set) of
initial states are quantum. Since measuring $\mathcal{W}_{\Omega
mn}$ requires the information about state populations only and can
be performed with invasive observations, implementing
$\mathcal{W}_{\Omega mn}$ can be more practical than implementing
$\mathcal{W}_{QQ}$~(\ref{wqq}) and $\mathcal{W}_{Q}$~(\ref{wq}).

\subsection{Examples}
To illustrate the effectiveness of our witnesses we now
present five example systems where they could be applied.  For
each example we choose which ever witness is more appropriate,
given the properties of that system.

\begin{figure}
\includegraphics[width=8.5cm]{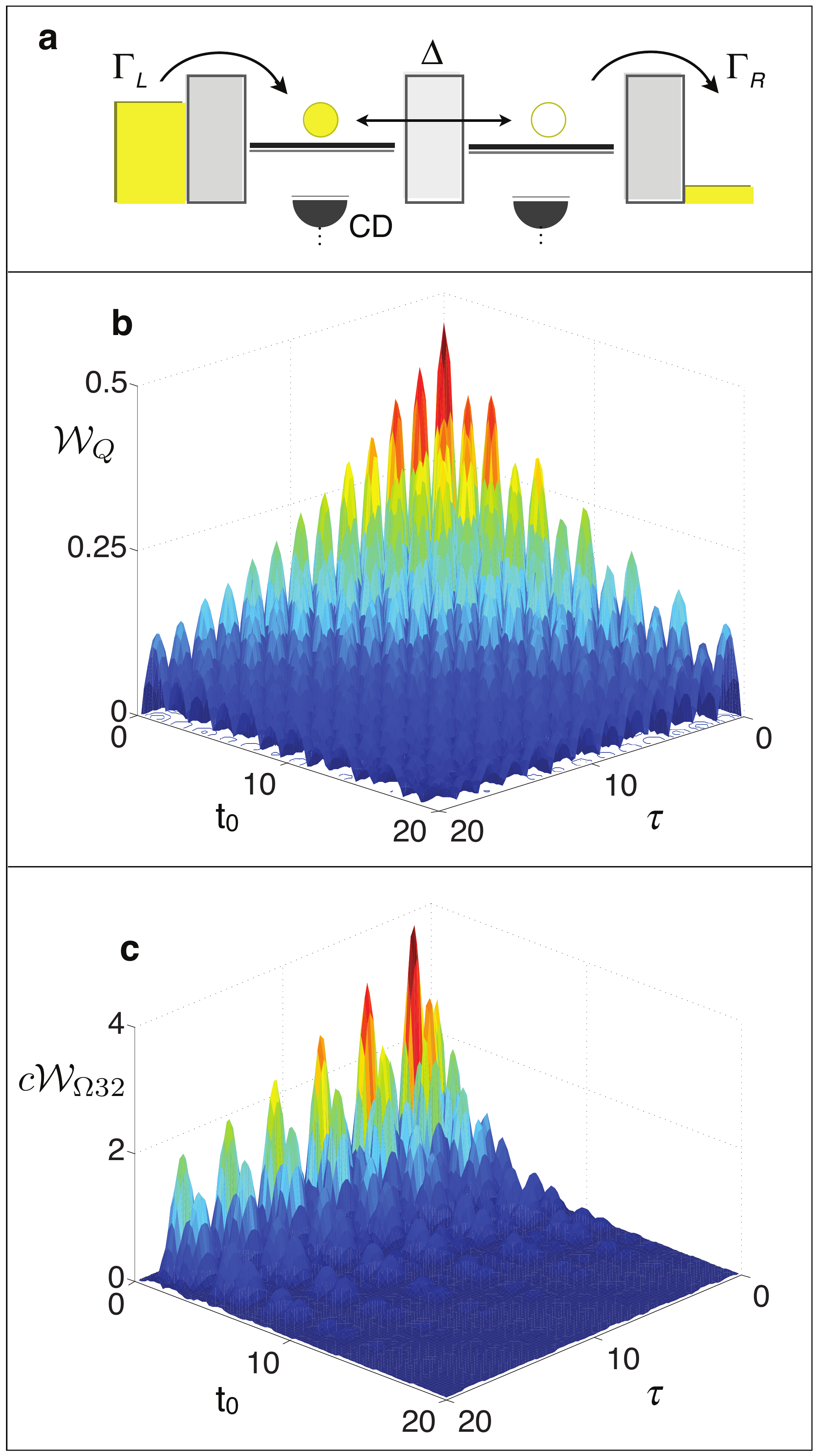}
\caption{\textbf{Detecting quantum transport through a double
quantum dot.} \textbf{a}, Schematic of a single-electron double
quantum dot (DQD). Here we assume that the DQD is weakly coupled
to leads under a large bias. Its Hamiltonian is
$H_{\mathrm{\!D\!Q\!D}}=\Delta(\left|L\right\rangle\!\!\left\langle
R\right|+\left|R\right\rangle\!\!\left\langle L\right|)$ with the
electron state basis
$\{\left|L\right\rangle\!,\!\left|R\right\rangle\!,\!\left|0\right\rangle\}$
where $\Delta$ is the tunnelling amplitude between the left-dot
and right-dot electron states
$\left|L\right\rangle,\left|R\right\rangle$. The transport between
dots and leads is described by the self-energy,
$\Sigma[\rho]=-1/2\sum_{\alpha=L,R}\Gamma_{\alpha}[s_{\alpha}s_{\alpha}^{\dag}\rho-2s_{\alpha}^{\dag}\rho
s_{\alpha}+\rho s_{\alpha}s_{\alpha}^{\dag}]$, where
$s_{L}=\left|0\right\rangle\!\!\left\langle L\right|$,
$s_{R}=\left|R\right\rangle\!\!\left\langle 0\right|$, and
$\Gamma_{L}$ and $\Gamma_{R}$ are the left and right tunnelling
rates, respectively. We assume charge detectors (CDs) are used for
the measurements, but invasive current measurements are also
sufficient (not shown here). \textbf{b},\textbf{c}, Verifying
quantum transport through DQD with $\mathcal{W}_{Q}$
[Eq.~(\ref{wq}) for $t-t_{0}=\tau$] and $\mathcal{W}_{\Omega 32}$ [Eq.~(\ref{wo})],
respectively. Here we define $\left|0\right\rangle$,
$\left|L\right\rangle$, and $\left|R\right\rangle$ by
$\left|1\right\rangle$, $\left|2\right\rangle$, and
$\left|3\right\rangle$, respectively. For the setting
$\Gamma_{L}=4$, $\Gamma_{R}=0.1$, and $\Delta=1$, the non-vanished
$\mathcal{W}_{Q}$ and $c \mathcal{W}_{\Omega 32}$ indicate the
quantum-transport regions, where $c=(p_{1}p_{2}p_{3})^{-1}$ for
the stationary state.}
\end{figure}

\begin{figure}
\includegraphics[width=8.7cm]{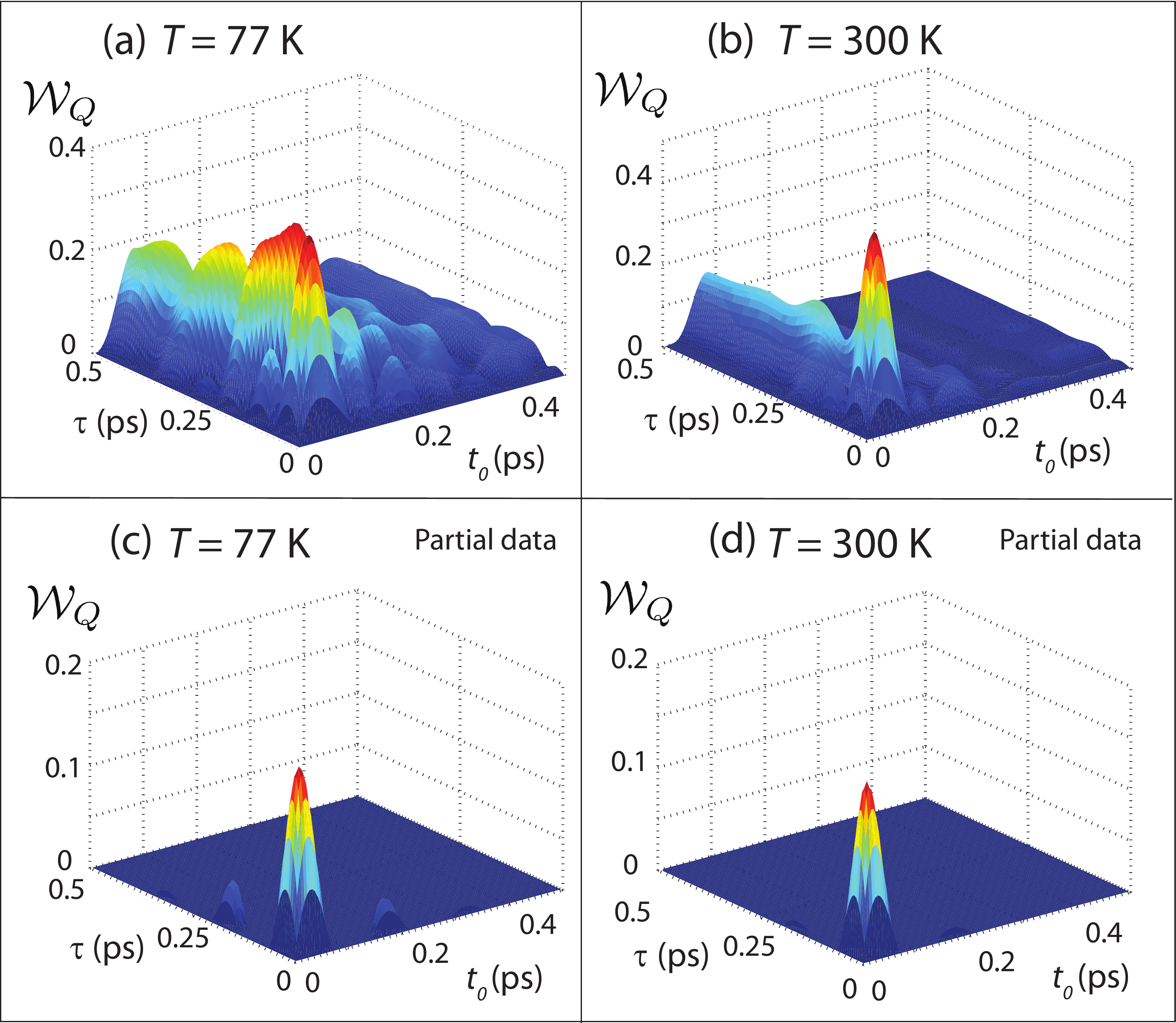}
\caption{\textbf{Detecting quantum properties of the FMO complex.}
Magnitude of the first witness $\mathcal{W}_{Q}$ [Eq.~(\ref{wq}) for $t-t_{0}=\tau$]
for the FMO complex assuming the final measurement is done on site
$m=1$ of the seven site FMO complex, for both \textbf{(a)} $T=77$
K and \textbf{(b)} $T=300$ K. A detection via our first witness
is clearly visible for an initial evolution greater than $t_0=0.3$ ps at $77$ K.  In
comparison, for same parameters we employ for the witness, the LG
inequality only reveals a violation for upto $0.035$ ps\cite{Wilde10}.
 \textbf{(c)} and \textbf{(d)} show the first witness with only limited
access, i.e. with only state preparation and measurement on site
$1$.  Quantum coherence is only detected when
$p_{1}(t_{0})\Omega_{11}(t,t_{0})> \left \langle
Q_{1}(t)\right\rangle$. In all figures the bath parameters used
were $\gamma^{-1} = 50$ fs and $\lambda=35$ cm$^{-1}$, and the
Hamiltonian is the same as that used in Ref. \cite{Ishizaki10}.
For the Hierarchy calculation, we used the ``Ishizaki-Tanimura''
truncation scheme and truncation as taken at $K=0$ and $N_c=8$
(see Methods, or Ref.~\cite{Ishizaki10}, for the meaning of these
parameters).}
\end{figure}

\subsection{Rabi oscillations in superconducting qubits} The
oscillations of state populations are commonly thought of as a
signature of quantum dynamics. The measurement of these kind of
oscillations is widely employed for many experiments. The
observation of such oscillations alone, however, is not definitive
evidence for the existence of quantum coherent dynamics and can
even be mimicked by the solutions of classical autonomous rate
equations, e.g., Ref.~\cite{Om08,Timm09}.

As a first example of the application of our witnesses we apply
$\mathcal{W}_{\Omega mn}$~(\ref{wo}) to  a two-level system
composed of the two lowest-energy states in a single-Cooper-pair
box\cite{Bouchiat98,you,Nakamura99}, Figure 2a.  We can take
$n=1$, $m=2$, for example together with the designation $T_{j}$:
$\{t_{0[j,k]}=(k+j-1)t_{0},t_{[1,k]}=(k+j-1)t_{0}+\tau|k=1,2\}$
for $j=1,2$, Figure 2b illustrates that the quantum witness
$\mathcal{W}_{\Omega 21}$ detects the presence of quantumness in
the Cooper-pair tunneling. Since only information about state
populations is required, this witness is easy to apply in practice
with simple invasive measurements and can be readily applied to
the existing experiments  in the time domain \cite{you,Nakamura99}
\emph{without} any additional experimental
overhead.

One can also consider an application of our witnesses to single-
and multiple-transmon qubits coupled to transmission lines in
circuit quantum electrodynamics \cite{Chow09,reed2}, where
qubit-state measurements are performed by monitoring the
transmission through the microwave cavity \cite{reed2}. For the
simplest case of one-qubit rotation, the coherent evolution is
driven by the Hamiltonian\cite{Chow09} \beq H=\hbar
\omega\left|1\right\rangle\!\!\left\langle
1\right|+\varepsilon(t)(\left|0\right\rangle\!\!\left\langle
1\right|+\left|1\right\rangle\!\!\left\langle 0\right|),\eeq where
$\varepsilon(t)$ is the microwave pulse to induce transitions
between qubit states $\left|0\right\rangle$ and
$\left|1\right\rangle$ with an energy difference $\hbar \omega$.
Through properly choosing the pulse $\varepsilon(t)$, a reliable
single-qubit gate, e.g., the Hadamard transformation ($\text{H}$),
can be created. Here, we use the quantum-process-tomography-based
optimal control theory \cite{Hwang12} to design the microwave pulse
for such a gate ($\mathcal{E}_{\text{H}}$) with a process fidelity
of about $94\%$. We use the first  witness $ \mathcal{W}_{Q}$ in the form: \beq
\mathcal{W}_{Q}:=\bigg|\left \langle
0(\mathcal{E}_{2\text{H}})\right\rangle-\sum_{n=0}^{1}p_{n}(\mathcal{E}_{\text{H}})\Omega_{0n}(\mathcal{E}_{\text{H}})\bigg|,\eeq
to show that the process $\mathcal{E}_{\text{H}} $ creates
coherent rotations. When setting the input state as $\left|0\right\rangle$, the value
of our witness is about $\mathcal{W}_{Q}\approx 0.45$, which
certifies the quantumness of $\mathcal{E}_{\text{H}}$.

\subsection{Quantum transport in quantum dots} Experimentally distinguishing
quantum from classical transport through nanostructure remains a
critical challenge in studying transport phenomena and designing
quantum electronic devices. As mentioned in the introduction,
using time-domain methods to verify quantum coherence, such as by
testing the Leggett-Garg inequality, can be very demanding. We
illustrate here how our witnesses are valid under
invasive measurements by modelling single-electron transport
through double quantum dots (Figure 3a). Compared with the time
periods identified by the Leggett-Garg-type
approach \cite{Lambert10}, the quantum witnesses $\mathcal{W}_{Q}$
(Figure 3b) and $\mathcal{W}_{\Omega mn}$ (Figure 3c) can detect a
much larger quantum coherence window. For $\mathcal{W}_{\Omega
mn}$, we employ the settings $T_{j}$:
$\{t_{0[j,k]}=[k+c'(j-1)]t_{0},t_{[1,k]}=[k+c'(j-1)]t_{0}+\tau|k=1,2,3\}$
for $j=1,2$. Here $c'$ is large such that the whole system is
stationary in $T_{2}$.

\subsection{Energy transfer in a light-harvesting complex} As an example of the
effect of strong interactions with a bath we use a model from
bio-physics; energy transport in the
 Fenna--Matthews--Olson (FMO) pigment-protein
complex, where there is thought to be significant system-bath
entanglement and coherence \cite{Ishizaki10}.  As mentioned
earlier, this example enables to discuss the issue of whether
classical-correlations between system and bath can cause a
violation of our first witness $\mathcal{W}_{Q}$ (the second witness is not valid in
this regime).

\begin{figure}
\includegraphics[width=8.5cm]{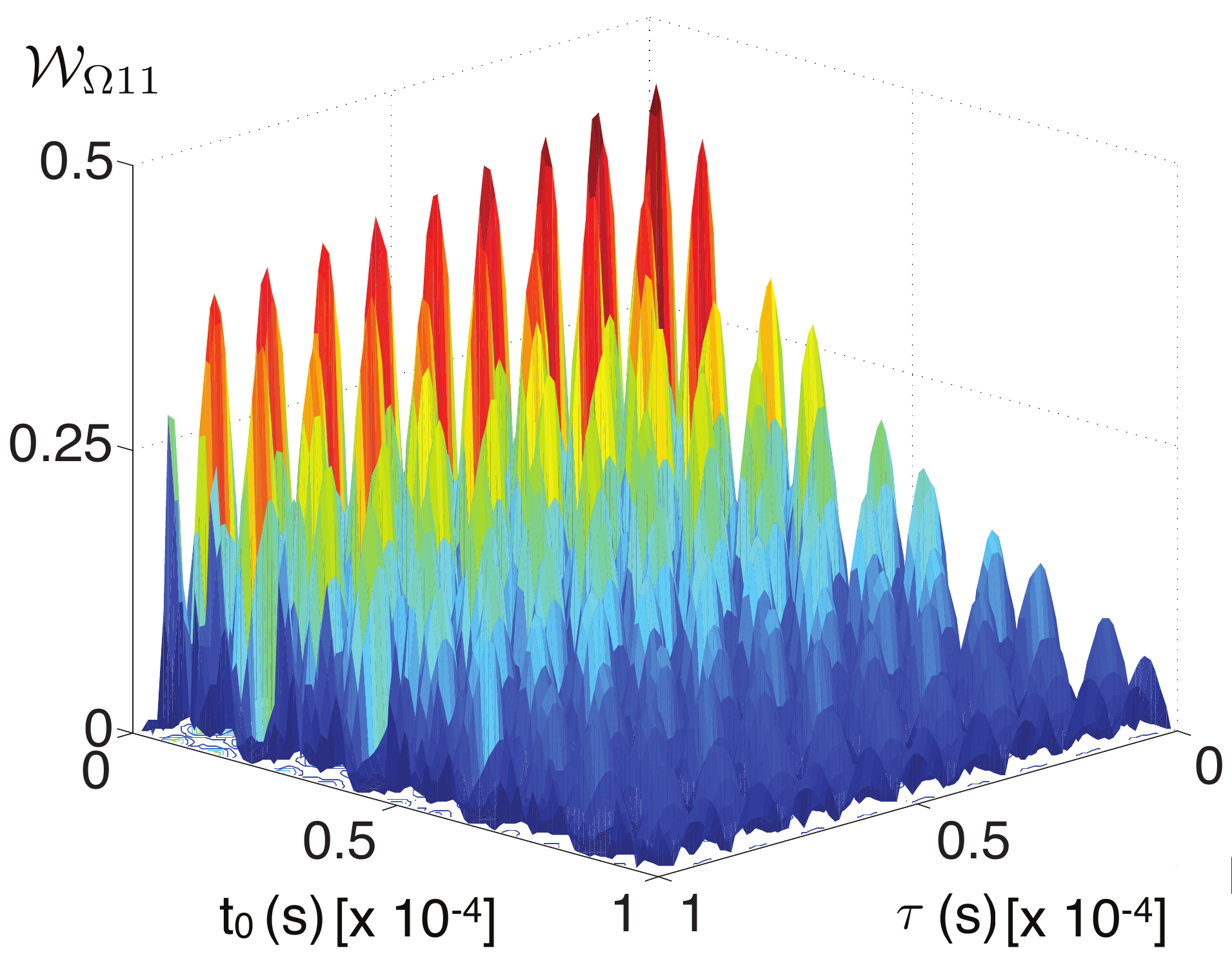}
\caption{\textbf{Detecting vacuum Rabi oscillations in a lossy
cavity.} Here we use the experimental data reported in
Ref. \cite{Brune01} to illustrate coherence-verification using our
second witness Eq.~(\ref{wo}). The circular Rydberg states with
principle quantum numbers $51$ and $50$ for transition
$\omega_{0}=51.1$ GHz are considered as the states
$\left|e\right\rangle$ and $\left|g\right\rangle$, respectively.
The atom-field coupling is $\omega_{R}/2\pi=47$ KHz. For a
high-$Q$ cavity with $Q=7 \times 10^{7}$, the vacuum Rabi
oscillation is detected by use of $\mathcal{W}_{\Omega mn}$ where
$m=n=1$. As a comparison we also checked the case when the
$Q$-factor is so low that $2\omega_{R}< \omega_{0}/Q$. For such a
low-$Q$ cavity (e.g., $Q=7 \times 10^{5}$), the state evolution is
in the regime of irreversible transitions and obeys the classical
constraint (3). Hence the value of the witness is zero.}
\end{figure}

In the methods section we impose a classical condition based on an
assumption of a class of classical states.  States which violate
this assumption possess coherences (either in the internal system
degrees of freedom, or in the system-bath degrees of freedom,
i.e., entanglement). However, to prevent classical correlations
between system and bath from causing a false positive, the
propagators $\Omega_{mn}(t,t_0)$ in our witness (\ref{wq}), which
we construct by preparing the system in one (or more) of its
states, must also capture the classical correlations between
system and reservoir present at time $t_0$. In the other examples
we discuss in this work, this is trivial since the system and bath
are always in a product state. However, in systems like the FMO
complex we discuss here, this is not the case.  Thus to account
for these correlations when constructing $\Omega_{mn}(t,t_0)$ in a
general case we must do the following: prepare the system-bath product state at $t
= 0$, evolve to time $t_0$, and perform a measurement on the system to
project it, without preserving coherence,  onto one of it states $n$.
 We then evolve again, retaining the post-measurement system-bath
state, and deduce the propagator by measuring the occupation of the
state $m$ at final time $t$.  If we can do ideal projective
(non-coherence preserving) measurements this accounts for the
classical system-bath correlation loophole (as long as we can
consistently prepare the $t=0$ separable system-bath state). If we are
doing destructive or invasive measurements then we must be able to
re-prepare the destroyed system state, at time $t_0$, on a time scale
faster than the bath/environment dynamics. Since there is no need for measurements on superpositions of basis states, this procedure can be performed without quantum tomography.

We illustrate this with the FMO complex, a seven-site structure
used by certain types of bacteria to transfer excitations from a
light-harvesting antenna to a reaction center. It has been the
focus of a great deal of attention due to experimental observation
of apparent ``quantum coherent oscillations'' at both $77$ K and
room temperature.   To fully capture the non-Markovian and
non-perturbative system-bath interactions of this complex system
we employ the Hierarchical equations of
motion \cite{Ishizaki09,Ishizaki10}, an exact model (given a bath
with a Drude spectral density) valid for both strong system-bath
coupling and long-bath memory time. We use the parameters used by
Ishizaki and Fleming in Refs. \cite{Ishizaki09,Ishizaki10}, and in
Figure 4 we show how this model is detected as quantum by our
witness $\mathcal{W}_{Q}$  , even at room temperature. We also show, in Figure
4c and 4d, how only partial information about the terms in
propagator is needed to find a detection at small times, thus
reducing the experimental overhead.  In constructing the
propagator terms for the sum in Eq.~(\ref{wq}) in this case we
discard all coherence terms in the physical density matrix but
retain the state of the bath, as in \cite{TanimuraPRL}.  In this
way we account for the state of the bath at time $t_0$, as
discussed above.  However, accounting for the classical
correlations with the reservoir seems beyond the capability of
current experiments.  We also point out that the full witness detects
coherence on timescales greater than $t_0=0.3$ ps at $77$ K, which is a much larger detection
window than the Leggett-Garg inequality ($0.035$ ps) for the same
parameters \cite{Wilde10}.

\begin{figure}
\includegraphics[width=8.5cm]{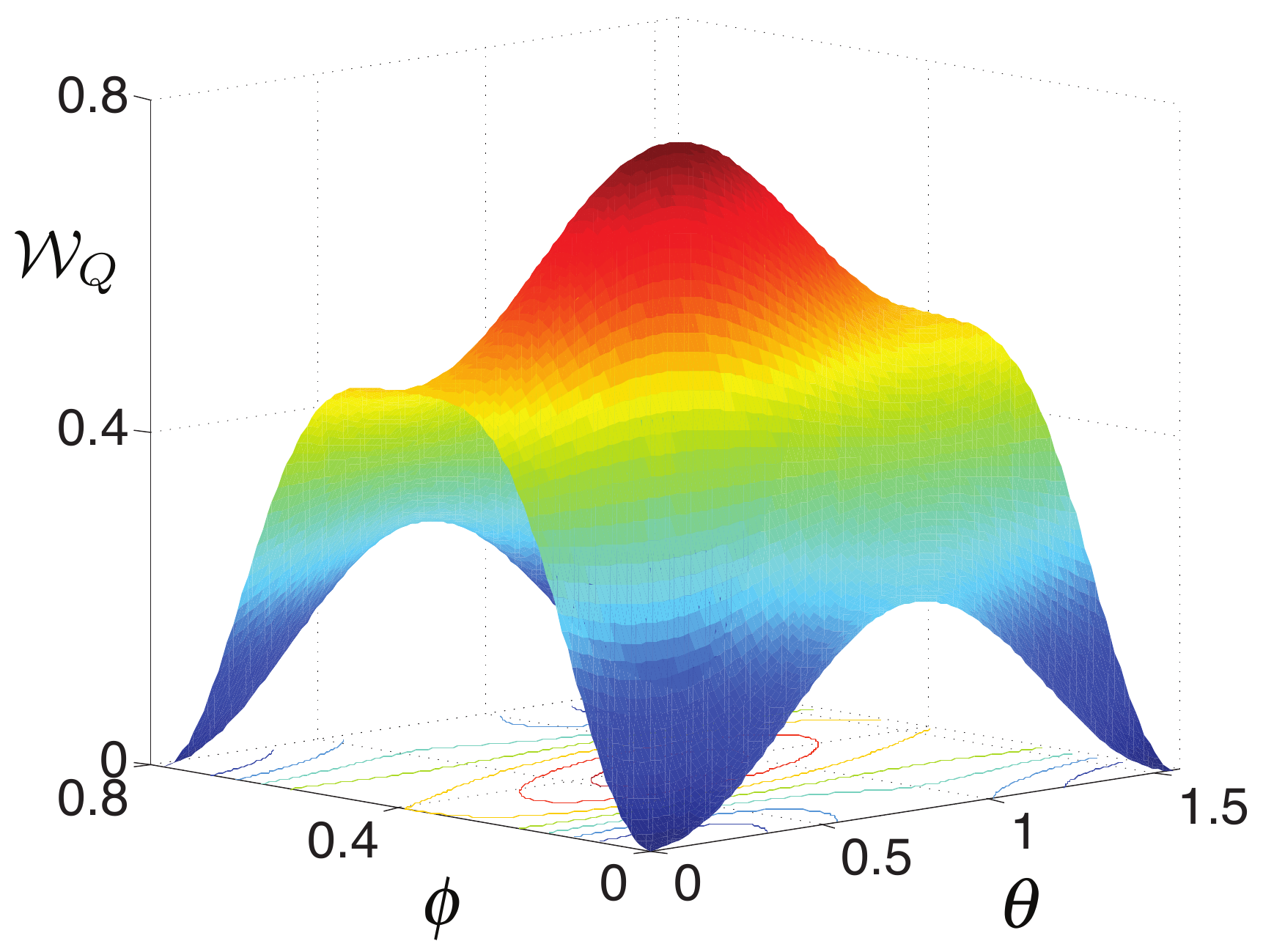}
\caption{\textbf{Detecting coherent rotations of photonic qubits.}
The first witness $\mathcal{W}_{Q}$, Eq.~(\ref{wq}), adapted from
the time-domain to the domain of the angles of several applied
transformations, detects quantum coherence in almost the whole
range of the prepared states $\rho_0$ (see text) as a function of
different angle settings of wave plates $(\phi,\theta)$:
$0\leq\phi\leq \pi/4$ and $0\leq\theta\leq \pi/2$.}
\end{figure}

\subsection{Vacuum Rabi oscillation in a lossy cavity} We now
consider a Rydberg atom placed in a single-mode cavity which is in
resonance with an atomic transition frequency, $\omega_{0}$, for
an adjacent pair of circular Rydberg states \cite{Raimond01}
$\left|e\right\rangle$ and $\left|g\right\rangle$. Let us know
consider the case when the cavity field are initially prepared in
the excited state $\left|e\right\rangle$ and the vacuum state
$\left|0\right\rangle_{p}$, respectively (denoted by
$\left|1\right\rangle=\left|e\right\rangle\left|0\right\rangle_{p}$).
In this case, the atom-field state becomes
$\left|2\right\rangle=\left|g\right\rangle\left|1\right\rangle_{p}$
due to spontaneous emission and then periodically oscillates
between the states $\left|e\right\rangle\left|0\right\rangle_{p}$
and $\left|g\right\rangle\left|1\right\rangle_{p}$ at the vacuum
Rabi frequency $\omega_{R}$. If the field irreversibly decays due
to photon loss out of the cavity, the atom-field stochastically
evolves to
$\left|3\right\rangle=\left|g\right\rangle\left|0\right\rangle_{p}$
from $\left|2\right\rangle$. Summarizing the above, the time
evolution of the atom-field state $\rho$ can be described by the
following master equation\cite{Gerry05}
\begin{equation}
\frac{d}{dt}\rho =
-\frac{i}{\hbar}[H_{\mathrm{JC}},\rho]-\frac{\kappa}{2}(\hat{a}^{\dag}\hat{a}\rho+\rho
\hat{a}^{\dag}\hat{a})+\kappa\hat{a}\rho\hat{a}^{\dag}
\end{equation}
where $H_{\mathrm{JC}}=\frac{\hbar
\omega_{R}}{2}(\hat{a}\sigma_{+}+\hat{a}^{\dag}\sigma_{-})$ is the
interaction Hamiltonian of the system. Here $\kappa=\omega_{0}/Q$,
and $Q$ is the quality factor of the cavity.

We now use our second witness to detect the vacuum-Rabi oscillation
between the atom and cavity field states. Here we choose the
time-domain set as $T_{j}$:
$\{t_{0[j,k]}=(k+j-1)t_{0},t_{[1,k]}=(k+j-1)t_{0}+\tau|k=1,2,3\}$
for $j=1,2$. Figure 5 shows the value of the witness for
vacuum-Rabi oscillations in a high-$Q$ cavity. Using the
experimental parameters from \cite{Brune01}, where $2\omega_{R}\gg
\omega_{0}/Q$, the damped coherent oscillations of the atom-cavity
state are detected as quantum by our second witness, shown in Fig.
5a. In comparison, for a low-$Q$ cavity, where $2\omega_{R}<
\omega_{0}/Q$, irreversible spontaneous emission out of the cavity
will dominate the state evolution. The value of the witness
$\mathcal{W}_{\Omega mn}$ is zero for this case. The measurements
on atom states we require to construct the witness are
experimentally available by using field-ionization
detectors \cite{Raimond01} for selecting atom states
$\left|e\right\rangle$ and $\left|g\right\rangle$.

\subsection{Coherent rotations of photonic quantum bits} Photon
polarization states $\left|H\right\rangle$ (horizontal) and
$\left|V\right\rangle$ (vertical) have been widely used to achieve
linear optical quantum information processing, quantum
communication, and quantum metrology \cite{Pan08,Kok07,OBrien09}.
As a qubit, polarization states can be coherently manipulated by
half-wave plates (HWP) and quarter-wave plates (QWP). Arbitrary
qubit rotations can be performed by using these linear optics
elements. Here we will use our first quantum witness
$\mathcal{W}_{Q}$ to detect the quantum coherence of polarization
states created by these rotations. The transformations of HWP and
QWP can be represented by the following\cite{James01}:
\begin{eqnarray}
&&H_{wp}(\phi)=\cos(2\phi)(\left |H\right\rangle\!\!\left\langle H\right|-\left |V\right\rangle\!\!\left\langle V\right|)\nonumber\\
&&\ \ \ \ \ \ \ \ \ \ \ \ \ \ \ \ \ \ \ \ \ \ \ \ \ \ -\sin(2\phi)(\left |H\right\rangle\!\!\left\langle V\right|+\left |V\right\rangle\!\!\left\langle H\right|),\\
&&Q_{wp}(\theta)=\frac{1}{\sqrt{2}}\big[(i-\cos(2\theta))\left |H\right\rangle\!\!\left\langle H\right|+(i+\cos(2\theta))\left |V\right\rangle\!\!\left\langle V\right| \nonumber\\
&&\ \ \ \ \ \ \ \ \ \ \ \ \ \ \ \ \ \ \ \ \ \ \ \ \ \
+\sin(2\theta)(\left |H\right\rangle\!\!\left\langle
V\right|+\left |V\right\rangle\!\!\left\langle H\right|\big)]
\end{eqnarray}
As a concrete example, one can set a HWP at $\phi=\pi/8$ to create
a photonic Hadamard gate $H_{wp}(\pi/8)$.

To detect the coherent rotations created by $R(\phi,
\theta)=Q_{wp}(\theta)H_{wp}(\phi)$, we use the first quantum
witness to probe the coherence between states
$\left|H\right\rangle$ and $\left|V\right\rangle$. While the
witness is originally constructed in the time domain, it can be
rephrased in terms of the settings $(\phi, \theta)$. Assuming that
both the wave plates are perfect and there is no photon loss in
the birefringent crystals of the wave plates, we have the
following correspondences: \beq \left\langle Q_{m}(\phi,
\theta)\right\rangle=\text{tr}[\left|m\right\rangle\!\!\left\langle
m\right|R(\phi, \theta)\rho_{0} R^{\dag}(\phi, \theta)],\eeq and
\beq \Omega_{mn}(\phi, \theta)=|\left\langle m\right|R(\phi,
\theta)\left|n\right\rangle|^{2},\eeq where $\rho_{0}$ is some
initial state created by $R$. Here $m=H$ and $n=V$ denote the
different measurement basis for the horizontal and vertical
polarizations. In this example, we set the initial state as
$\rho_{0}=R^{\dag}(\phi,
\theta)\left|m\right\rangle\!\!\left\langle m\right|R(\phi,
\theta)$ and then the witness becomes
\begin{eqnarray}
&&\mathcal{W}_{Q}=\bigg|1-\frac{1}{16}\big[10+2\cos(4\theta)+2\cos(4\theta-8\phi)\nonumber\\
&& \ \ \ \ \ \ \ \ \ \ \ \ \ \ \ \ \ \ \ \ \ +\cos(8\theta-8\phi)+\cos(8\phi)\big]\bigg|.
\end{eqnarray}
Figure 6 shows this quantum witness for different prepared states
$\rho_{0}$, as a function of the angles $\theta$ and $\phi$.

The usual approach to \emph{strictly} probe the coherent
superposition of states $\left|H\right\rangle$ and
$\left|V\right\rangle$ is via quantum state
tomography \cite{James01}. Compared to such tomographic
measurements on single qubit states, which require three local
measurement settings, only one setting of a local measurement is
now sufficient to implement our first witness.

\section*{Discussion}

In summary, we have formulated a set of quantum witnesses that
allow the efficient detection of quantum coherence, without the restriction of non-invasive
measurements. Compared to some of the existing methods, such as
the Leggett-Garg inequality or employing general quantum
tomography, our approach can drastically reduces the overhead and
complexity of unambiguous experimental detection of quantum
phenomena, and has a larger detection window. As illustrated by the five physical examples, these witnesses are robust and can be
readily used to explore the presence of quantum coherence in a
wide-range of complex systems, e.g., transport in nanostructures,
biological systems, and perhaps even large-arrays of qubits used
in adiabatic quantum computing \cite{DWAVE}. After this paper went to press, we became aware of this preprint \cite{Kofler12}, which has related results.

\section*{Acknowledgement}
We are grateful to Y. Ota  and C. Emary for helpful comments. C.-M.L. acknowledges the partial support from the National Science Council, Taiwan, under Grant No. NSC 101-2112-M-006-016-MY3, No. NSC 101-2738-M-006-005, and No. NSC 103-2911-I-006 -301, and the National Center for Theoretical Sciences (south). Y.-N.C. is partially supported by the National Science Council, Taiwan, under Grant No. NSC 101-2628-M-006-003-MY3. FN is partially supported by the ARO, JSPS-RFBR contract No. 12-02-92100, Grant-in-Aid for Scientific Research (S), MEXT Kakenhi on Quantum Cybernetics, and the JSPS via its FIRST program.

\section*{Methods}

\noindent \textbf{Proof of equation (1).} The quantum two-time state-state correlation
$\left\langle Q_{m}(t)Q_{n}(t_{0})\right\rangle_{Q}$ is defined
by \cite{Carmichael03}:
\begin{eqnarray}
&&\left\langle Q_{m}(t)Q_{n}(t_{0})\right\rangle_{Q}\nonumber\\
&=&\text{tr}_{S\!R}[ \rho_{S\!R}(0)Q_{m}(t)Q_{n}(t_{0})]\nonumber\\
&=& \text{tr}_{S}\big\{
Q_{m}(0)\text{tr}_{R}[U(\tau)\rho_{S\!R}(t_{0})Q_{n}(0)U^{\dag}(\tau)]\big\},
\end{eqnarray}
where $\rho_{S\!R}(t_{0})$ is the system-reservoir state and
$U(\tau)$ is the system-reservoir evolution operator for
$\tau=t-t_{0}$. If $\rho_{S\!R}(t_{0})$ is a classical state with
no coherent components, then we have
\begin{equation}
\rho_{S\!R}(t_{0})Q_{n}(0)=p_{n}(t_{0})Q_{n}(0)R(t_{0})
\end{equation}
 where $p_{n}(t_{0})$ is the probability of measuring the system state $n$ at time $t_0$ for the classical mixture $\rho_{S\!R}(t_{0})$, and $R(t_{0})$ is the reservoir state at time $t_{0}$ (which in principle depends on the measurement result $Q_n$ if the
 system and reservoir are classically correlated, i.e., are separable but in a mixture of product states). Then we have
 \begin{eqnarray}
 &&\left\langle Q_{m}(t)Q_{n}(t_{0})\right\rangle_{Q}\nonumber\\
 &=& p_{n}(t_{0})\text{tr}_{S}\big\{ Q_{m}(0)\text{tr}_{R}[U(\tau)Q_{n}(0)R(t_{0})U^{\dag}(\tau)]\big\}.\nonumber
 \end{eqnarray}
  The term describing the system's evolution $\text{tr}_{R}[U(\tau)Q_{n}(0)R(t_{0})U^{\dag}(\tau)]$ can be described by the operator-sum representation\cite{Breuer07,Nielsen00}:
  \begin{equation}
  \text{tr}_{R}[U(\tau)Q_{n}(0)R(t)U^{\dag}(\tau)]=\sum_{j}E_{j}(\tau)Q_{n}(0)E_{j}^{\dag}(\tau),\nonumber
  \end{equation}
  where $E_{j}(\tau)=\sum_{k}\sqrt{p_{rk}}\left\langle r_{j}\right|U(\tau)\left|r_{k}\right\rangle$. The the reservoir state is assumed to be $R(t_{0})=\sum_{k}p_{rk}\left| r_{k}\right\rangle\!\!\left\langle r_{k}\right|$. Hence the correlation $\left\langle Q_{m}(t)Q_{n}(t_{0})\right\rangle_{Q}$ for the system-reservoir classical mixture at the time $t_{0}$ is
 \begin{eqnarray}
&&\left\langle Q_{m}(t)Q_{n}(t_{0})\right\rangle_{Q}\nonumber\\
&=&  p_{n}(t_{0})\text{tr}_{S}\big\{ Q_{m}(0)\sum_{j}E_{j}(\tau)Q_{n}(0)E_{j}^{\dag}(\tau)\big\}\nonumber\\
&=&  p_{n}(t_{0})\sum_{j}\Omega^{(j)}_{mn}( t,t_{0})\nonumber\\
&=&p_{n}(t_{0})\Omega_{mn}(t,t_{0}),
\end{eqnarray}
where $\Omega_{mn}(t,t_{0}):=\sum_{j}\Omega^{(j)}_{mn}(t,t_{0})$ is
the propagator, i.e., the probability of finding the state $m$ at the
time $t$ when the state at an earlier time $t_{0}$ is initialized at
$n$.\\

\noindent \textbf{The Hierarchy model for FMO.} The Hierarchy model was originally developed by Tanimura and
Kubo \cite{Tanimura3}, and has been applied extensively to
light-harvesting complexes \cite{Ishizaki09,Ishizaki10}. We will
not give a full description here, but will just summarize the main
equation and parameters. It is always assumed that at $t=0$ the
system and bath are separable $\rho(0)=\rho_S(0)\otimes
\rho_B(0)$, and that the bath is in a thermal equilibrium state
$\rho_B(0) = e^{-\beta H^{(B)}}/\mathrm{Tr}\left[e^{-\beta
H^{(B)}}\right]$, $\beta = 1/K_{\mathrm{B}}T$.  The bath is
 assumed to have a Drude spectral density \beq J_j(\omega) =
\left(\frac{2\lambda_j \gamma_j}{\hbar}\right)
\frac{\omega}{\omega^2 +\gamma_j^2}, \eeq where $\gamma_j$ is the
``Drude decay constant'' and each site $j$ is assumed to have its
own independent bath.  In addition, $\lambda_j$ is the
reorganisation energy, and is proportional to the system-bath
coupling strength. The correlation function for the bath is then
given by, \beq C_j = \sum_{m=0}^{\infty} c_{j,m}
\exp\left(-\mu_{j,m} t\right) \eeq where $\mu_{j,0} = \gamma_j$,
and $\mu_{j,m} = 2\pi m/\hbar \beta$ when $m\geq 1$. The
coefficients are \beq c_{j,0} = \gamma_j \lambda_j\left(\cot(\beta
\hbar \gamma_j/2) - i\right)/\hbar\eeq and \beq c_{j,m\geq 1} =
\frac{4\lambda_j \gamma_j}{\beta \hbar^2}
\frac{\mu_{j,m}}{\mu_{j,m}^2 -\gamma_j^2}. \eeq

Under these assumptions, the Hierarchy equations of motion are
given by, \beq \dot{\rho}_{\mathbf{n}} &=& -\large(i L +
\sum_{j=1}^N\sum_{m=0}^K \mathbf{n}_{j,m}
\mu_m\large)\rho_{\mathbf{n}} -
i\sum_{j=1}^N\sum_{m=0}^K\left[Q_j,\rho_{\mathbf{n}_{j,m}^+}\right]\nonumber\\
&-& i\sum_{j=1}^N\sum_{m=0}^K
n_{j,m}\left(c_mQ_j\rho_{\mathbf{n}_{j,m}^-} - c_m^*
\rho_{\mathbf{n}_{j,m}^-}Q_j\right). \eeq The operator
$Q_j=\ket{j}\bra{j}$ is the projector on the site $j$, and for FMO
there are seven sites, thus $N=7$. The Liouvillian $L$ describes
the Hamiltonian evolution of the FMO complex. The label
$\mathbf{n}$ is a set of non-negative integers uniquely specifying
each equation; $\mathbf{n}=\{n_1,n_2,n_3,...,n_N\} =
\{\{n_{10},n_{11},..,n_{1K}\},..,\{n_{N0},n_{N1},..,n_{NK}\}\}$.
The density matrix labelled by $\mathbf{n}=0=\{\{0,0,0....\}\}$
refers to the system density matrix, and all others are
non-physical density matrices,  termed ``auxiliary density
matrices''.  The density matrices in the equation labelled by
$\mathbf{n}_{j,m}^{\pm}$ indicate that that density matrix is the
one defined by increasing or decreasing the integer in the label
$\mathbf{n}$, at the position defined by $j$ and $m$, by $1$.

 The hierarchy equations must be truncated, which is typically done by
 truncating the largest total number of
terms in a label $N_c = \sum_{j,m} n_{j,m}$.  This value is termed
the tier of the hierarchy. The choice of $N_c$ should be
determined by checking the convergence of the system dynamics.
Here we also use the ``Ishizaki-Tanimura boundary
condition''\cite{Tanimura}; \beq L_{\mathrm{IT-BC}} =-
\sum_{j=1}^N\sum_{m=K+1}^{\infty} \frac{c_{j,m}}{\mu_{j,m}}
\left[Q_j,\left[Q_j,\rho_{\mathbf{n}}\right]\right].\eeq This can
be summed analytical, which for $K=0$ gives, \beq
\sum_{m=1}^{\infty} \frac{c_{j,m}}{\mu_{j,m}} =
\frac{4\lambda_j}{\hbar^2 \gamma_j
\beta}\left\{1-\gamma_j\hbar\left[\cot(\gamma_j\hbar
\beta/2)\right]\beta/2\right\}.\eeq

\end{document}